\documentclass[oneside,english,pre]{revtex4}
\usepackage{graphicx}
\usepackage{amsfonts, color}
\usepackage{amsthm}

\begin{document}

\title{Perturbation propagation in random and evolved Boolean networks}
\author{Christoph Fretter$^1$, Agnes Szejka$^2$ and Barbara Drossel$^2$}
\address{$^1$ Instistut f\"ur Informatik, Martin-Luther-Universit\"at Halle-Wittenberg, Von-Seckendorffplatz 1, 06120 Halle, Germany}
\address{$^2$ Institut f\"ur Festk\"orperphysik, TU Darmstadt, Hochschulstrasse 6, 64289 Darmstadt, Germany}
\email{Christoph.Fretter@informatik.uni-halle.de}

\begin{abstract} 
We investigate the propagation of perturbations in Boolean networks by
evaluating the Derrida plot and modifications of it. We show that even
small Random Boolean Networks agree well with the predictions of the
annealed approximation, but non-random networks show a very different
behaviour.  We focus on networks that were evolved for high dynamical
robustness.  The most important conclusion is that the simple
distinction between frozen, critical and chaotic networks is no longer
useful, since such evolved networks can display properties of all
three types of networks. Furthermore, we evaluate a simplified
empirical network and show how its specific state space properties
are reflected in the modified Derrida plots.
\end{abstract}
\maketitle

\section{Introduction}

Boolean networks (BNs) are used to model the dynamics of a wide variety
of complex systems, ranging from neural networks \cite{rosen-zvi} and
social systems \cite{moreira04} to gene regulation
networks \cite{lagomarsino}, sometimes combined with evolutionary
processes \cite{bassler04, agnespaper}. BNs are composed of interacting
nodes with binary states, typically $0$ and $1$, coupled among each
other.  The state of each node evolves according to a function of the
states observed in a certain neighbourhood, similar to what is done
when using cellular automata \cite{herrmann88}, but in contrast to
cellular automata, BNs have no regular lattice structure, and not all
nodes are assigned the same update function. 

An often studied class of BNs are Random BNs (RBNs). In such networks,
connections between nodes and functions are assigned at random. The
dynamics of RBNs can be classified into frozen, chaotic and critical
\cite{derrida}. In a frozen network, a perturbation at one node
propagates in one time step on an average to less than one other node.
Starting from a random initial state, such networks freeze after a short
time completely (apart from possibly a small number of nonfrozen
nodes). Therefore attractors of the dynamics have typically length 1
(i.e., they are fixed points) or are very short. In a chaotic
network, a perturbation at one node propagates in one time step on an
average to more than one other node.  The attractors of the dynamics
are long, and a nonvanishing proportion of all nodes remains nonfrozen.
Critical networks are at the boundary between these two phases, a
perturbation at one node propagates in one time step on an average to
exactly one other node.

Whether a RBN is frozen, critical or chaotic, can be determined by
using the annealed approximation, which is a mean-field theory. It
neglects correlations between nodes or states of nodes and corresponds
to a system where the connections are newly assigned at each time
step. Furthermore, since the annealed approximation neglects
fluctuations, it applies to the limit of infinite network size.
Usually, the phase diagram derived analytically by using the annealed
approximation is correct for RBNs.

RBNs were originally introduced by S. Kauffman as a simple model for
gene regulation networks \cite{kauffman:metabolic}. Since he
considered frozen and chaotic networks as biologically unrealistic, he
suggested that biological networks are in a critical state ``at the
edge of chaos''. Although it has been shown in the meantime that
critical RBNs have unrealistically large mean attractor numbers and
attractor lengths \cite{samuelssontroein,ourscalingpaper}, the idea
that cellular networks may be poised at a critical state is still
alive. The concept of phase transitions, borrowed from equilibrium
statistical physics, is thus carried over to biological systems, which
are far from equilibrium. 

In this paper, we address the question whether the simple
classification of the dynamics of networks into frozen, chaotic and
critical is still appropriate when networks are not assembled
randomly. In fact, it was found that networks that were generated by
some evolutionary algorithm, may have a very long but unique
attractor that attracts all of state space \cite{sevim}, or that in
such networks initially similar states may diverge at first
exponentially fast but then converge to the same fixed point
\cite{agnespaper}. These are two examples where features of the
``frozen'' and ``chaotic'' phases are united within the same network.

In order to better understand when and how some networks can combine
features of different phases, we investigate the so-called Derrida
plot \cite{derrida}. The Derrida plot maps the initial Hamming
distance between the states of two identical networks (the number of
nodes with differing states) onto that obtained after one time step.
We will show in the next section that for RBNs the properties of the
Derrida plot and modifications derived from it agree with the predictions
of the annealed approximation and with the simple classification into
frozen, critical and chaotic networks. In section \ref{evolved}, we
focus on networks that were evolved for high dynamical robustness. We
will see that the annealed approximation cannot be applied
consistently to such networks, and that features of different
``phases'' are visible in the Derrida plot and related graphs.
In section \ref{yeastcycle}, we evaluate the Derrida plot and its
modifications for a model of the yeast cell cycle network \cite{Li},
which shows again deviations from the annealed approximation.

\section{RBNs}
\label{RBN}

\subsection{The Derrida plot}

We consider RBNs of $N$ nodes with a fixed number $K$ of randomly
assigned inputs per node and with the update function at each node
chosen at random among all possible update functions. Such networks
are frozen for $K<2$, critical for $K=2$, and chaotic for $K>2$ \cite{annealed}.

In order to evaluate the Derrida plot, two copies of the same network
are initialized in the same random initial state. Then the state of
$h$ nodes of one copy are inverted. The Hamming distance between the
two copies is now $h$. Then, both networks are updated once (i.e., one
time step is considered), and the Hamming distance is evaluated again.
For a given network, this procedure is repeated several times for each
value of $h$ from 0 to $N$, and the average is taken for each value
$h$.

\begin{figure}[t]
	\hspace{2.4cm} \includegraphics[width=0.35  \textwidth]{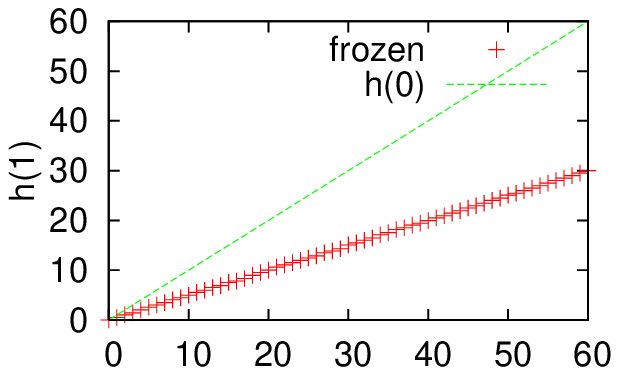}
	\includegraphics[width=0.35 \textwidth]{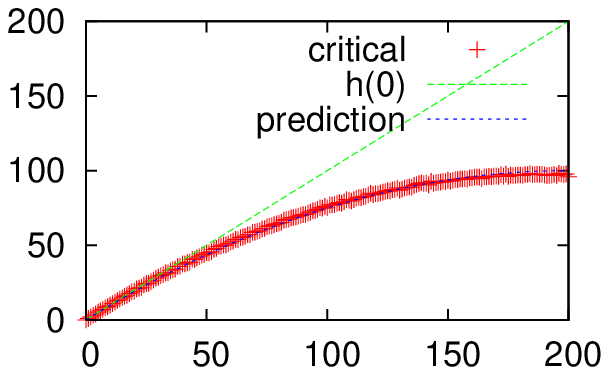}

	\hspace{2.4cm} \includegraphics[width=0.35 \textwidth]{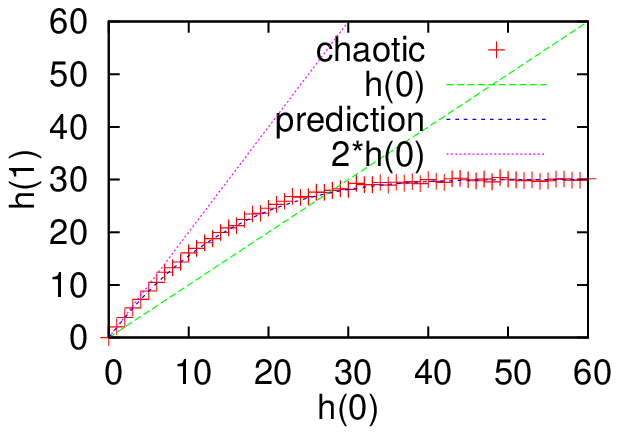}
	\includegraphics[width=0.35 \textwidth]{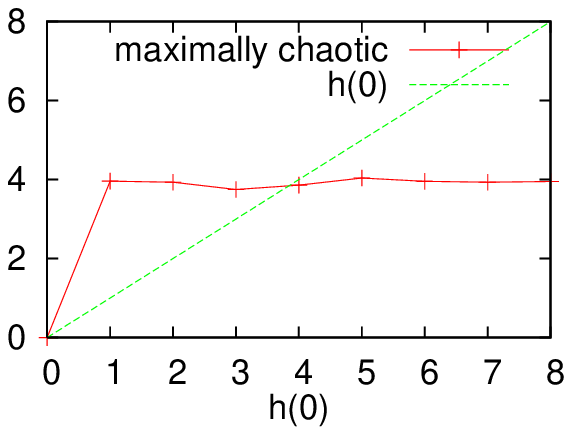}
\caption{Derrida plot of RBNs with $K=1$, $N=60$; $K=2$, $N=200$; $K=4$, $N=60$; and $K=N=8$. The lines connecting the symbols  are only guides to the eye.}
\label{fig1}
\end{figure}

The resulting Derrida plots are shown in figure \ref{fig1} for a
frozen ($K=1$, $N=60$), a critical ($K=2$, $N=200$), a chaotic ($K=4$,
$N=60$) and a maximally chaotic ($K=N$ for $N=8$) network. Each point
in the four plots is the average over at least 200 different
perturbations and initial conditions.

The Derrida plot has the following properties, which can be easily
explained within the annealed approximation \cite{annealed,Kesseli}:
\begin{enumerate}
\item In the frozen phase ($K=1$), the Derrida plot is linear and has
  the slope 1/2. In networks with $K=1$, there are 4 possible update
  functions, two of which are constant functions, the other two being the ``copy'' function
  and the ``invert'' function. These are assigned to the nodes with
  equal probability, which means that half of the nodes have a
  constant function and are therefore at time $t=1$ certainly in the
  same state in the two copies. The probability that a given node is
  not in the same state in the two copies at time $t=1$ is therefore
  0.5 times the probability $h/N$ that the state of its input node
  differs at time $t=0$ in the two copies.
\item  For $K=2$, the initial slope of the Derrida plot has the
  critical value 1. The probability that the state of a given node is
  different in the two copies at time $t=1$ is 
\begin{eqnarray}
\frac {h(1)}{N}= r (1-(1-h(0)/N)^K)\, , \label{fundamental} 
\end{eqnarray}
which is the probability that the state of at least one of its inputs is
different in the two copies, times the probability $r$ that a difference
in one or more inputs leads to a difference in the output. 
The slope of the function $h(1)$ vs $h(0)$ is $rK$ for small $h(0)$. When using
all update functions with the same probability (as we have done), $r$
is equal to $1/2$.
\item In the chaotic phase, with fixed $K$, and with $N \gg K$, the initial
  slope of the Derrida plot is $Kr=K/2$.  The probability that the state of
  a given node is different in the two copies at time $t=1$ is again
  given by equation (\ref{fundamental}) with $r=1/2$. 
\item For RBNs with $K=N$, the average distance between any two
  non-identical states is $N/2$ after one time step, because each node
  changes its state with probability $r=1/2$, since the inputs differ in at
  least one node \cite{derACKreview}.
\end{enumerate}

Although the simulated networks are relatively small, their Derrida
plot agrees very well with the one predicted by the annealed
approximation. The relation between $h(1)$ and $h(0)$ derived from equation (\ref{fundamental}) matches perfectly the
results of the simulations for the critical and the chaotic network,
so that it is hard to distinguish the two lines in figure \ref{fig1}.

In order to prepare the discussion in section \ref{evolved}, it is
useful to consider also other graphs related to the Derrida plot,
which will then be used to compare random and evolved networks.

\subsection{The generalized Derrida plot}

\begin{figure}[t]

	\hspace{2.4cm}\includegraphics[width=0.55\textwidth]{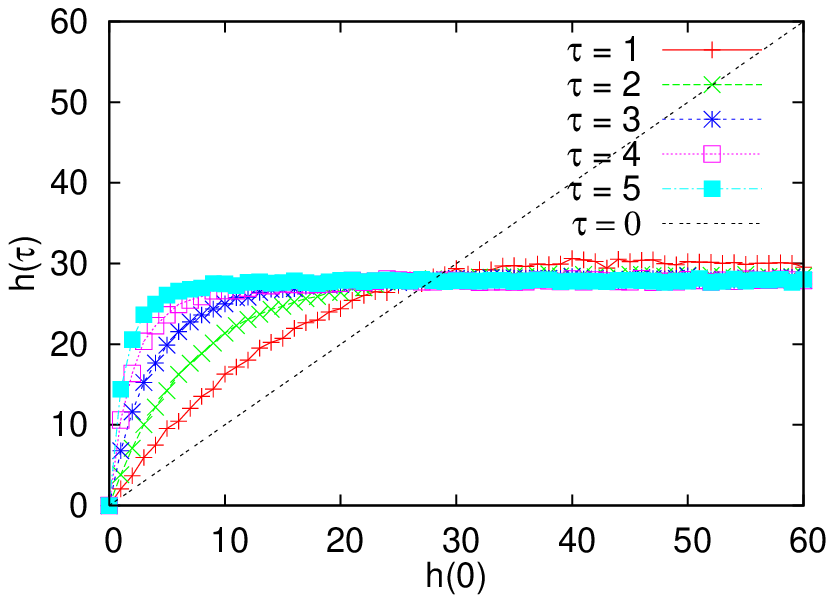}
	\caption{Generalized Derrida plot of a network with $K=4$, $N=60$, for different numbers of iterations $\tau$.}
	\label{fig2}

	\hspace{2.4cm}\includegraphics[width=0.55\textwidth]{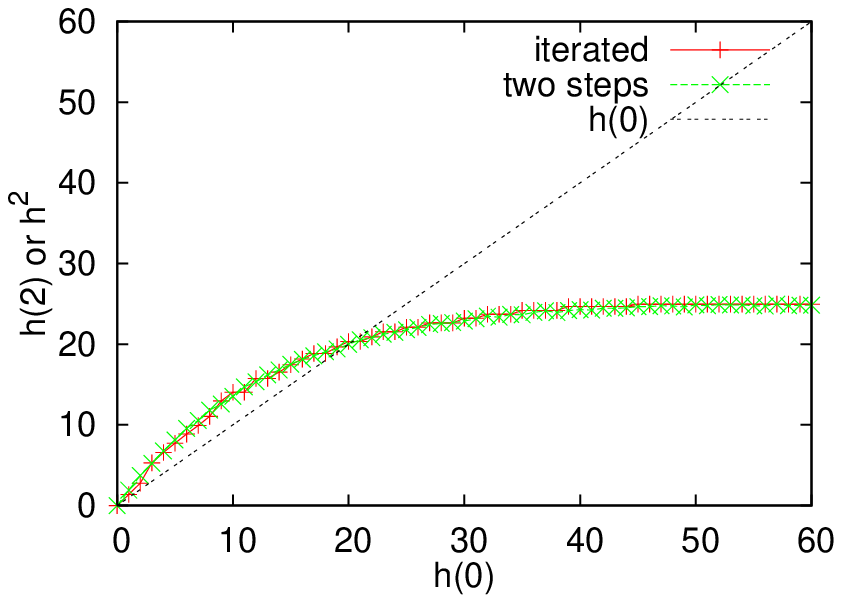}
	\caption{Generalized Derrida plot $h(2)$ (``two steps'', x
          symbols) and iterated Derrida map $h^2$ (``iterated'', +
          symbols) of a network with $K=3$, $N=60$.}
	\label{randMap}
\end{figure}

First, we generated generalized Derrida plots by evaluating the
Hamming distance $h(t>1)$ after more than one iteration as function of
$h(0)$. We will compare the generalized Derrida plot with the iterated
Derrida map. If we denote the Derrida map by $h(1)=f(h(0))$, the
iterated map is $f(f(h(0)))$ for two iterations, and correspondingly
for higher iterations.  In the following, we will denote the twice
iterated map $f(f(h(0)))$ simply by $h^2$.

Figure \ref{fig2} shows $h(t)$ as function of $h(0)$ for $t=1,...,5$
for the above chaotic network with $K=4$. Each data point in this and
in the following figures is averaged over 1000 randomly chosen
combinations of the initial state and the perturbed nodes.  The lines
connecting the points are only guides to the eye. Within the annealed
approximation, we expect $h(2)= h^2$, and equivalently for larger
times.  This is because the annealed approximation neglects
correlations between nodes or between different times, and therefore
the Hamming distance at one moment in time depends only on the Hamming
distance in the previous time step, but not on the precise state of
the system.  If the annealed approximation is valid for the
generalized Derrida plot for the RBN, not only the relation $h(2)=h^2$
must be satisfied, but also all plots for $h(t>1)$ must have the same
fixed points, i.e.  they must intersect the bisector at the same
points.  Figures \ref{fig2} and \ref{randMap} show that these two
conditions are indeed satisfied. For critical and frozen networks, we
found the same quality of agreement.  Further below, we will see that
no such agreement can be seen for evolved networks.

Such a good agreement between the annealed approximation and the
simulation data is even more surprising considering the fact that the
sizes of the networks are relatively small. The annealed
approximation can be expected to become exact in the thermodynamic
limit of infinite system size.

\subsection{The modified Derrida plot}

\begin{figure}[t]

	\hspace{2.4cm}\includegraphics[width= 0.55\textwidth]{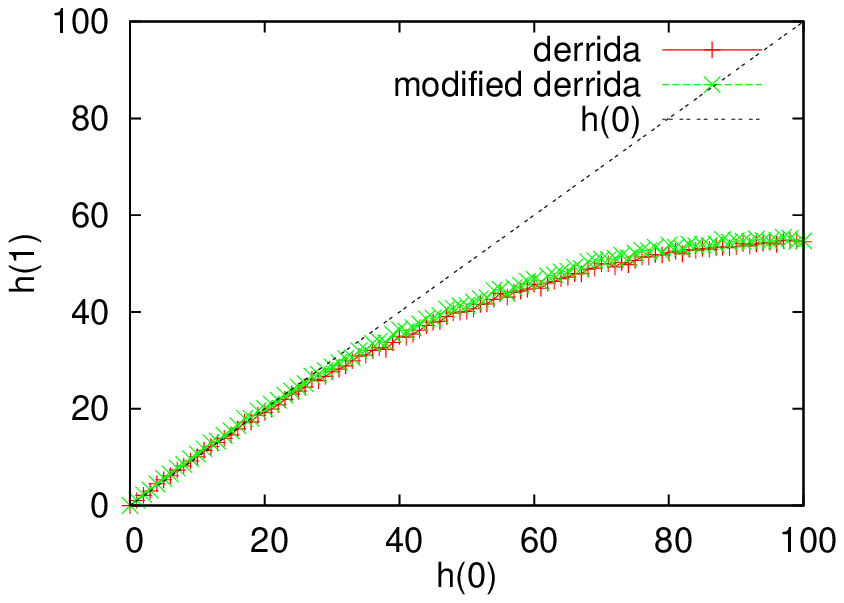}
	\caption{Derrida plot (+) and modified Derrida plot (x) of a
          network with $K=2$, $N=100$.}
	\label{fig3}
\end{figure}

Finally, we checked the validity of the annealed approximation by
comparing the Derrida plot obtained when starting from random initial
states with the Derrida plot obtained when starting only from states
that are on an attractor. In order to obtain initial states that are
on an attractor, we started from a random state and ran the simulation
until an attractor was reached. In this way, attractors are weighted
by the size of their basin of attraction.

Within the annealed approximation, there is
no concept of an attractor. An attractor is a periodic sequence of states,
and it can only occur if the connections between nodes are fixed in
time, and not annealed. Since an attractor contains only a small
subset of all states, one might expect that the Derrida plot becomes
different when the initial state is constrained to be on an attractor.
We therefore evaluated a modified Derrida plot, where the initial
state was an attractor state.  Figure \ref{fig3} shows the Derrida
plot and the modified Derrida plot for a critical network of size
$N=100$. The two curves agree very well with each other. We found an
equally good agreement for frozen and chaotic networks. This result
implies that states on attractors are ``typical'' states, which cannot
be distinguished from transient states when evaluating the temporal
behaviour of the Hamming distance.

\section{Networks evolved for robustness}
\label{evolved}
In this section, we investigate the Derrida plot and its modifications
for networks that were evolved for robustness of the dynamical
attractors against perturbations of the node values.  Like real
networks, they are not random but shaped by their evolutionary
history. We used networks evolved according to the rules described in
previous work by two of the authors \cite{agnespaper}, where the
evolution of a single Boolean network is simulated by means of an
adaptive walk.  The adaptive walk is a hill climbing process that
leads to a local fitness maximum and thus can yield insight into the
fitness landscape of a system.  In \cite{agnespaper}, it was found
that there is a huge plateau with the maximum fitness value that spans
the network configuration space.  Mutations change the connections and
the update functions of the nodes. In the following, we describe the
algorithm in more detail.

\subsection{Algorithm used for the network evolution}
The adaptive walk starts with a randomly created network. The fitness
(that is the robustness) of this network is determined by the
following procedure: First, the network is updated until it reaches an
attractor. Then the value of each node is flipped one after the other,
and it is counted how often the network returns to the same attractor
and how many time steps this takes. The fitness value is the percentage of times the dynamics
return to the given attractor after flipping a node, provided that the
way back to the attractor takes on average no longer than two
updates. If this condition is not fulfilled, the fitness is set to
zero.  After determining the fitness, a mutation is performed in the
network. With equal probability, a connection between two nodes is
added or deleted, a connection is redirected, or the Boolean function
is changed. The mutation is accepted if it does not lower the fitness.
The adaptive walk is continued until at least 15,000 accepted
mutations have occurred. Since the networks reach maximum fitness after
only a few evolutionary steps, most of these mutations are neutral
(i.e., they do not change the fitness value), and the networks
perform a random walk on the plateau of 100\% fitness. The
simulations are run for different $N$ up to $N = 90$ and initial
connectivities of $K_{ini} = 1$ and $2$. Here we show representative
networks with $N = 11$, $50$ and $80$. For details on the algorithm,
see \cite{agnespaper}. In contrast to most simulations performed in
that paper, we do not use canalyzing functions here, but the entire
set of Boolean functions, as in the RBNs evaluated in the previous
section.

\subsection{General properties of evolved networks}

We generated a set of evolved networks and evaluated first their
structural and dynamical properties, which are briefly summarized in
the following. The final networks have an average connectivity that
is larger than 2 and therefore characteristic for the chaotic regime.
The ensemble of evolved networks shows an average $K$ value of $3.1$ 
independently of the initial connectivity.
Unlike the initial random networks, which have the same value of $K$
for all nodes, the evolved networks have a broad distribution of input
connections, with a peak at 1 and an exponential decay for larger $K$.
The distribution of output connections has not changed and is
poissonian, as in random networks. The distribution of Boolean
functions in the evolved networks is the same as for the initial
networks, i.e., each function occurs with the same probability. We
also evaluated the ``sensitivity'' $\lambda$ of the evolved networks,
which is $K$ times the probability that a node changes its state when
one of its inputs is changed \cite{Shmulevich04}. It is identical to
the initial slope of the Derrida plot. We found $\lambda > 1$, which
also indicates that the final networks are in the chaotic regime.  The
large robustness of the evolved networks is due to a state space
containing an attractor with a huge basin that comprises usually more
than 90\% of all possible states.  Despite the high $K$ values, these
networks show only short attractor lengths, mostly single-digit, with
only a few active nodes on the attractors.

\subsection{Derrida plots of evolved networks}

\begin{figure}[t]

	\hspace{2.4cm}\includegraphics[width=0.55 \textwidth]{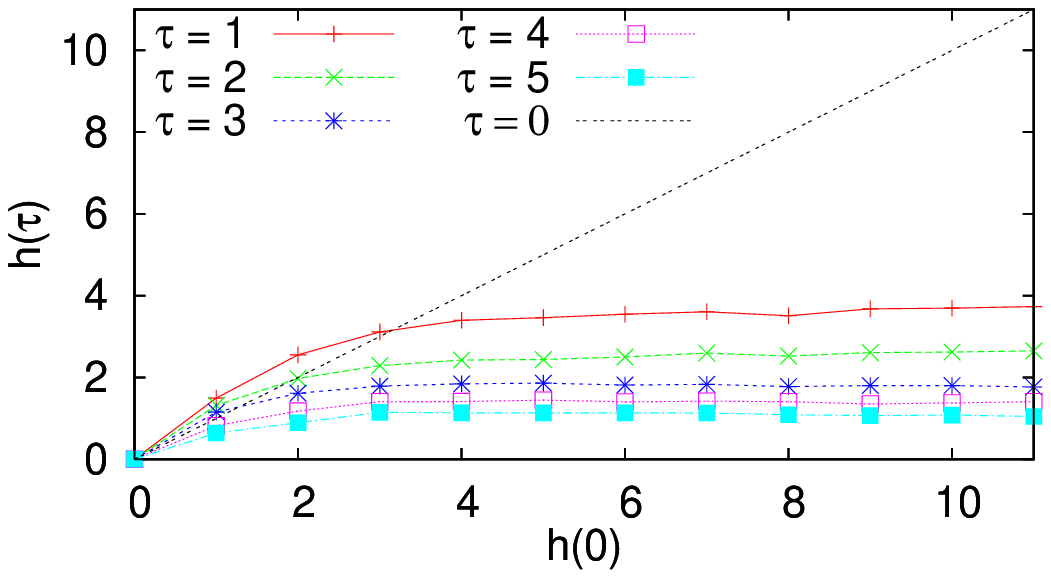}

	\hspace{2.4cm}\includegraphics[width= 0.55\textwidth]{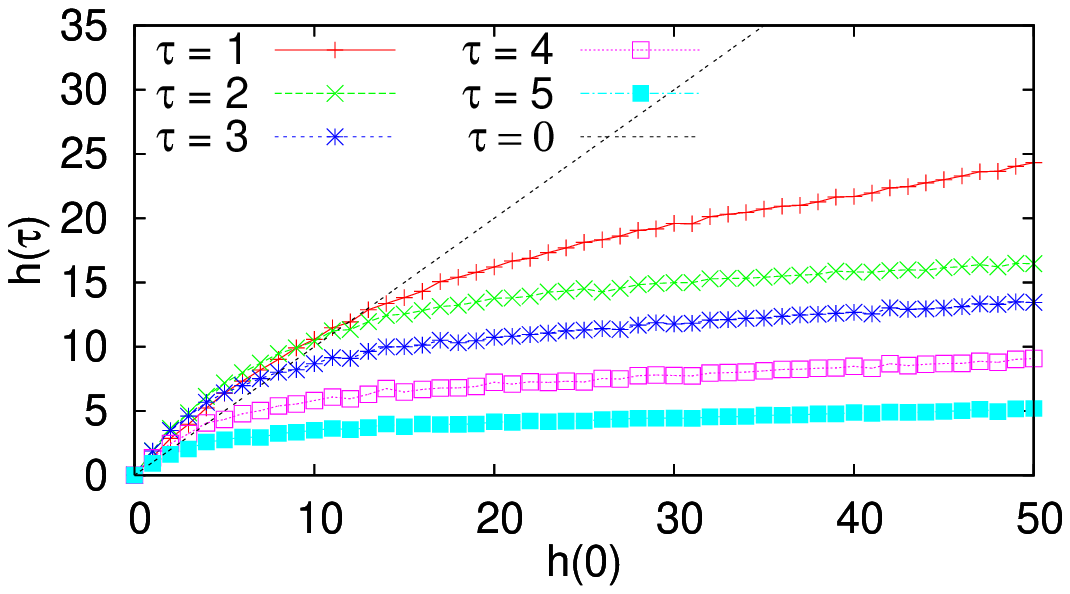}

	\hspace{2.4cm}\includegraphics[width= 0.55\textwidth]{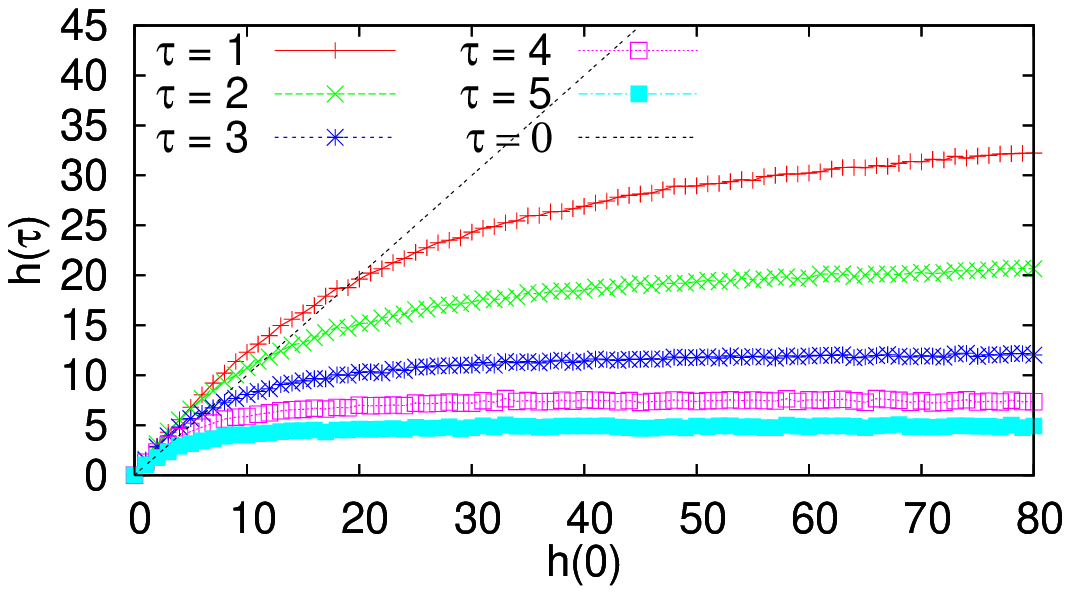}
	\caption{Generalized Derrida plots with $\tau$ = 1,2,3,4,5 of
	three networks that have been evolved for robustness. The
	parameters are  $N=11$, $K=3.63$; $N=50$, $K=3.12$; $N=80$, $K=3.39$.}
	\label{fig6}
\end{figure}

\begin{figure}[t]
	\hspace{2.4cm}\includegraphics[width= 0.55\textwidth]{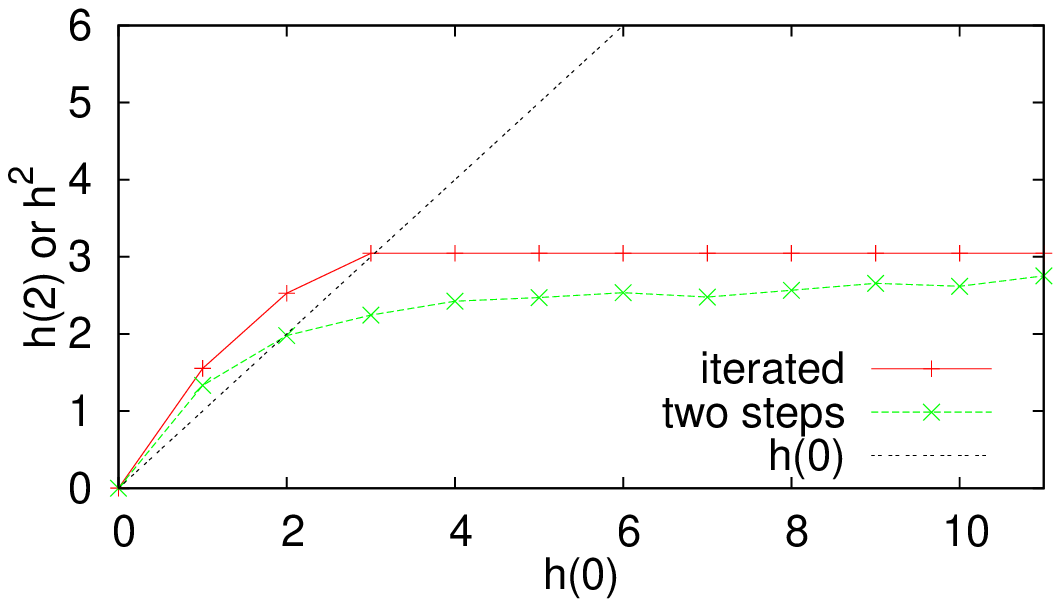}

	\hspace{2.4cm}\includegraphics[width= 0.55\textwidth]{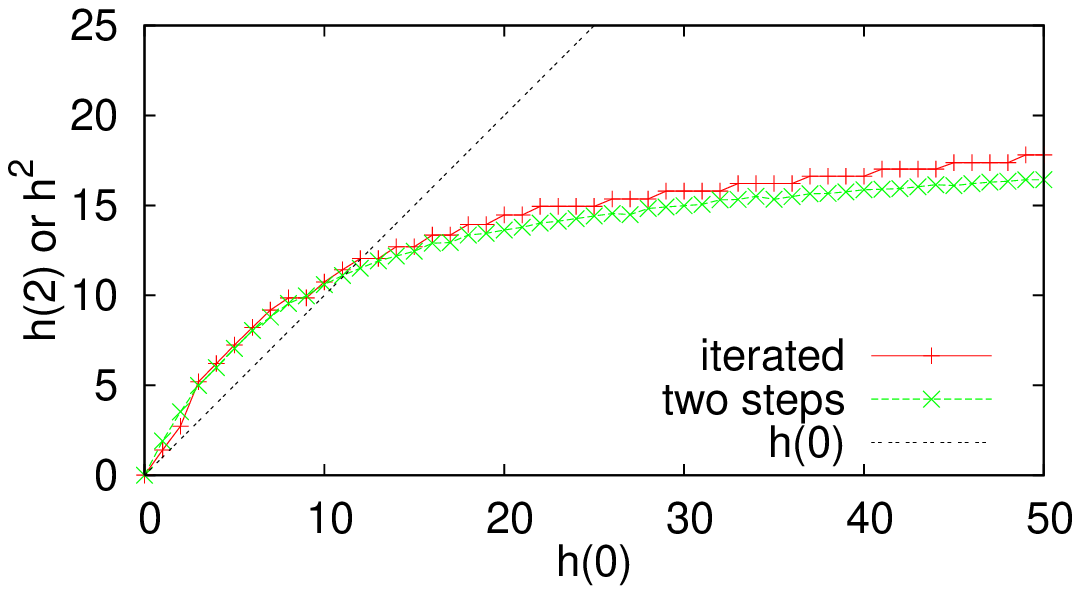}

	\hspace{2.4cm}\includegraphics[width= 0.55\textwidth]{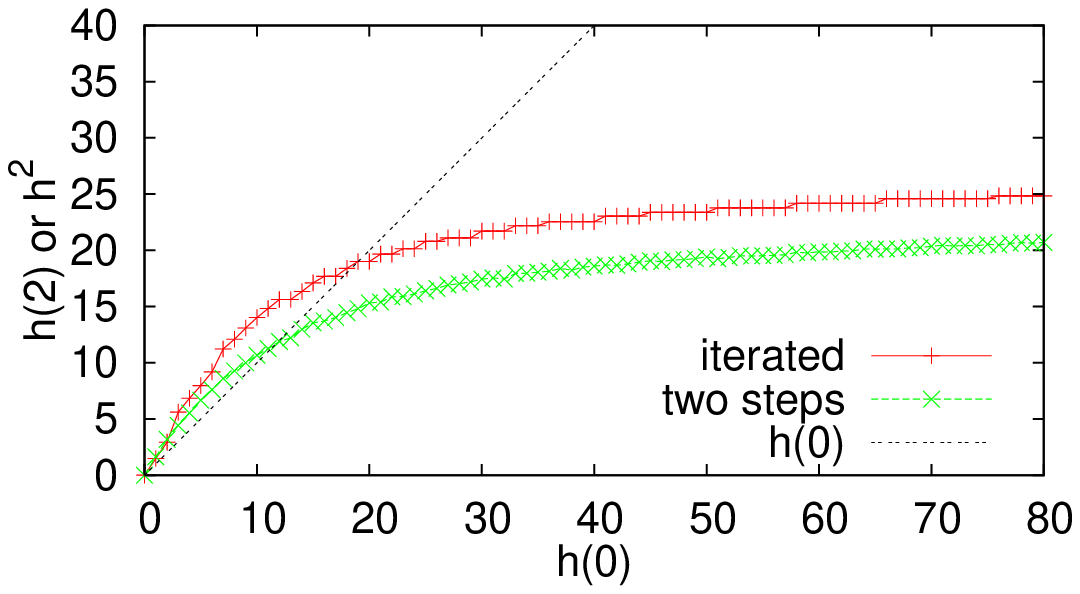}
	\caption {Generalized Derrida plot $h(2)$ (x symbols), and
        iterated Derrida map $h^2$ (+ symbols) of the three networks
        evolved for robustness $N=11$, $K=3.63$, $N=50$, $K=3.12$ and
        $N=80$, $K=3.39$.}
	 \label{fig4}
\end{figure}

\begin{figure}[t]
	\hspace{2.4cm}\includegraphics[width= 0.55\textwidth]{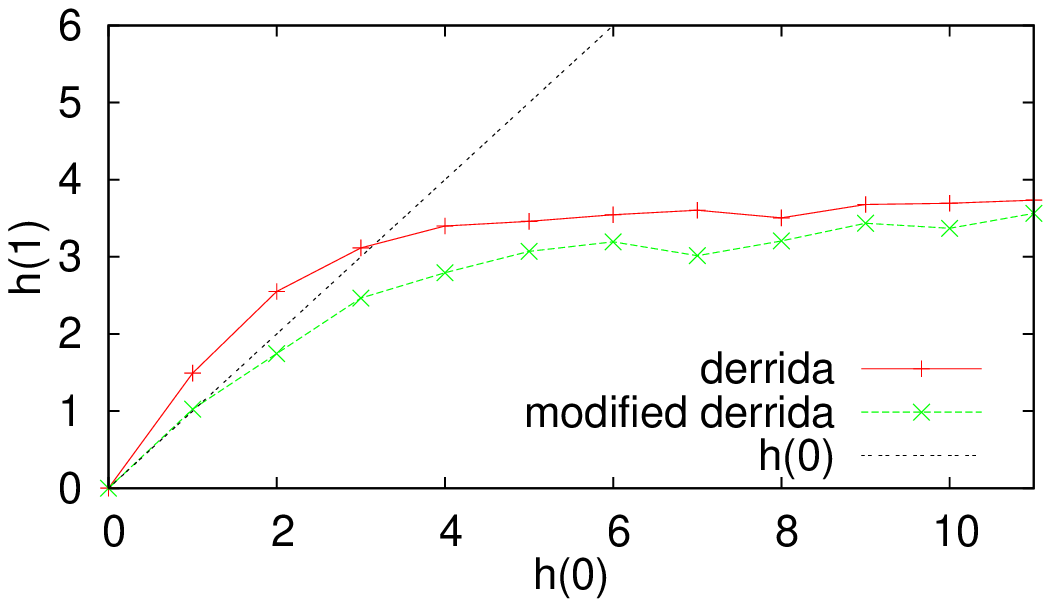}

	\hspace{2.4cm}\includegraphics[width= 0.55\textwidth]{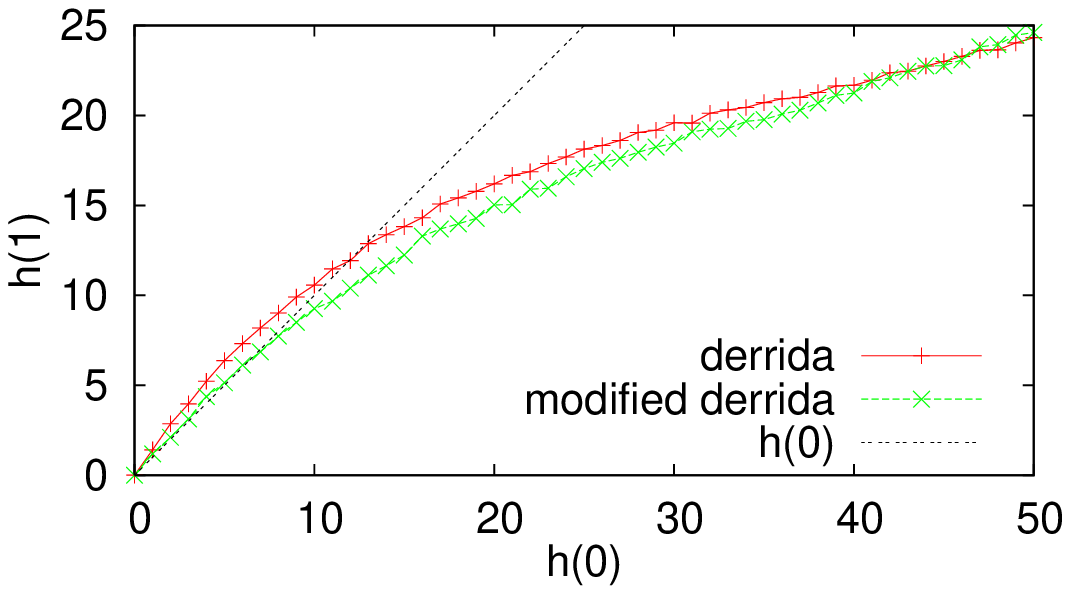}

	\hspace{2.4cm}\includegraphics[width= 0.55\textwidth]{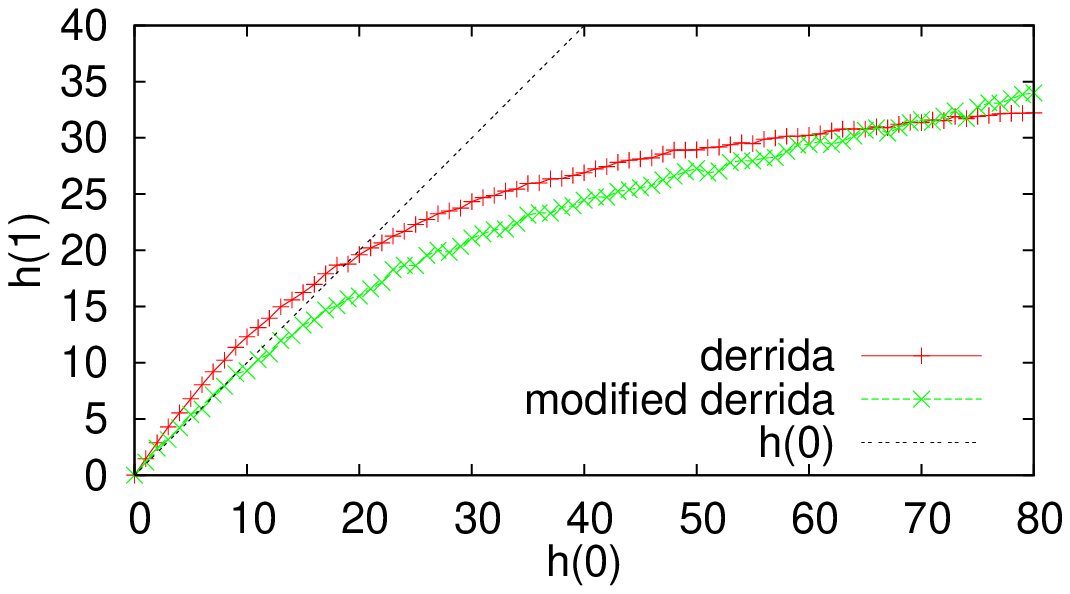}
	\caption{Derrida plot and modified Derrida plot of the three networks evolved for robustness $N=11$, $K=3.63$, $N=50$, $K=3.12$ and $N=80$, $K=3.39$.}
	\label{fig5}
\end{figure}

The Derrida plots and their generalized, iterated and modified versions are 
shown in figures \ref{fig6} to \ref{fig5} for three evolved networks
of different sizes.

Figure \ref{fig6} shows the generalized Derrida plots. They are
clearly different from those of RBNs. For RBNs, the curves intersect all
at the same point, which is a fixed point of the iterated Derrida
maps.  This fixed point value must be identical to 0.5 times the
number of nonfrozen nodes, $N_{nf}$, in the stationary state, since
each of these nodes change their state with probability 0.5 during one
iteration. Since the iteration must converge to this fixed point, the
generalized Derrida plots of RBNs approach with with increasing $\tau$
a step function with the value 0 at $h=0$ and with the fixed point value
$0.5 \cdot N_{nf}$ for all $h>0$ (see figure \ref{fig2}).  This is in
contrast to the evolved networks, where the maximum value reached by
the curves decreases with increasing $\tau$, and so does the value of
the fixed point of the map. Even though the initial states may be far
from an attractor, the trajectories in state space experience
constraints due to the fact that they are attracted to the same small
part of the state space. The plot shows how with increasing $\tau$
more and more nodes freeze while the system approaches the attractor.

Figure \ref{fig4} shows the curves of the two-step Derrida plot and
the iterated Derrida map for the same three networks. 
In contrast to RBNs (see figure \ref{randMap}), the corresponding
curves of evolved networks do not agree with each other. The two-step
Derrida plots are for larger $h$ values below the iterated Derrida
map. This indicates again that trajectories do not move away from each
other as fast as suggested by the first iteration. 

Finally, in figure \ref{fig5} we show the modified Derrida plot of the
evolved networks. The initial slope of these plots is close to 1,
which means that the networks appear to be critical for perturbations
occurring on attractors.  In contrast, the initial slope is larger
than 1 for the conventional Derrida plot, which means that the evolved
networks appear at first chaotic when starting from a random initial
state.  Such a decrease in the initial slope is plausible as these
networks were evolved for stable attractors.

We have shown here only the results for three networks. For other
evolved networks, the qualitative behaviour of the curves that we just
described is the same. The precise shape of the curves, however,
differs between different networks of the same size. 

All these results indicate that the state space structure of these
evolved networks is fundamentally different from that of RBNs. They
contain one large dominant attractor, which influences the Hamming
distance between trajectories already during the first few time steps.
The evolved networks combine features characteristic of the frozen
phase (the very short attractor and the very small number of
attractors), of the chaotic phase (the initial slope of the Derrida
plot), and of the critical point (the initial slope of the modified
Derrida plot, the large number of frozen nodes).

\section{The yeast cell cycle network}
\label{yeastcycle}

\begin{figure}[t]
	\hspace{2.4cm}(a) 

	\hspace{2.4cm}\includegraphics[width= 0.55\textwidth]{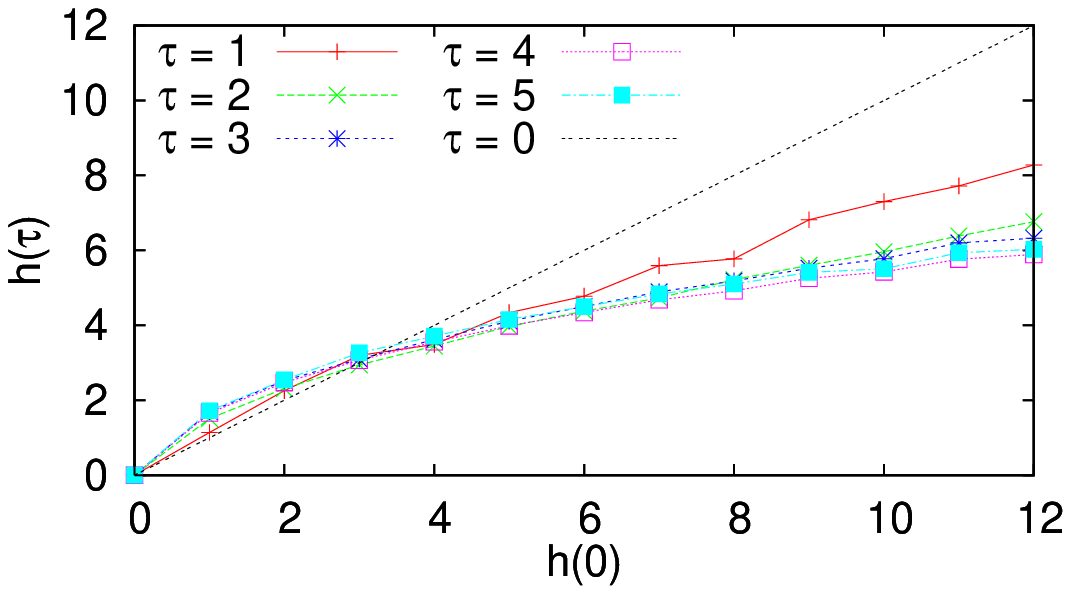}

	\hspace{2.4cm}(b)

	\hspace{2.4cm}\includegraphics[width= 0.55\textwidth]{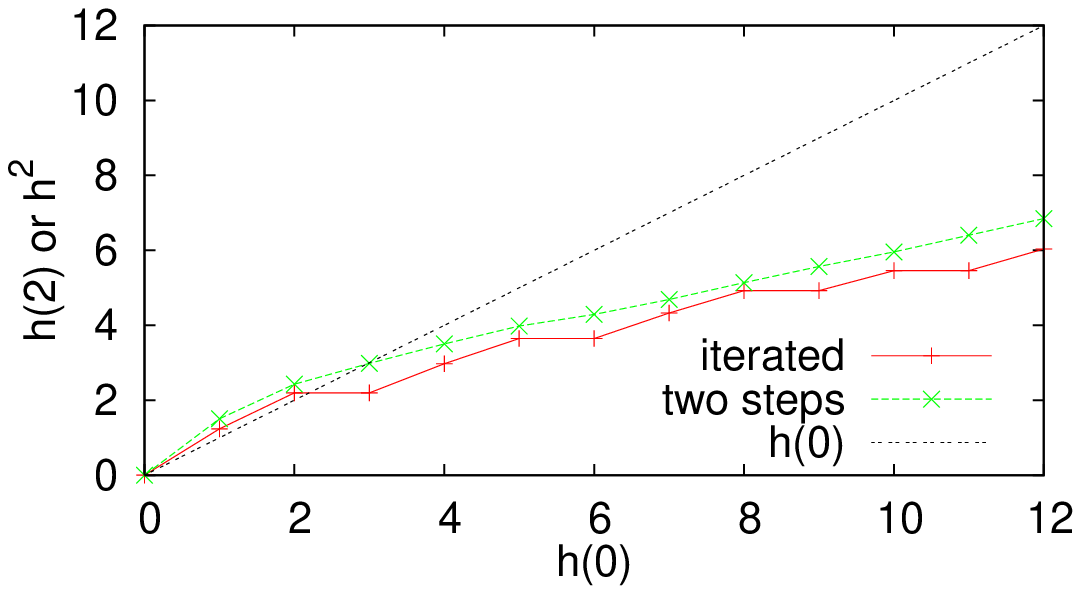}

	\hspace{2.4cm}(c)

	\hspace{2.5cm}\includegraphics[width= 0.55\textwidth]{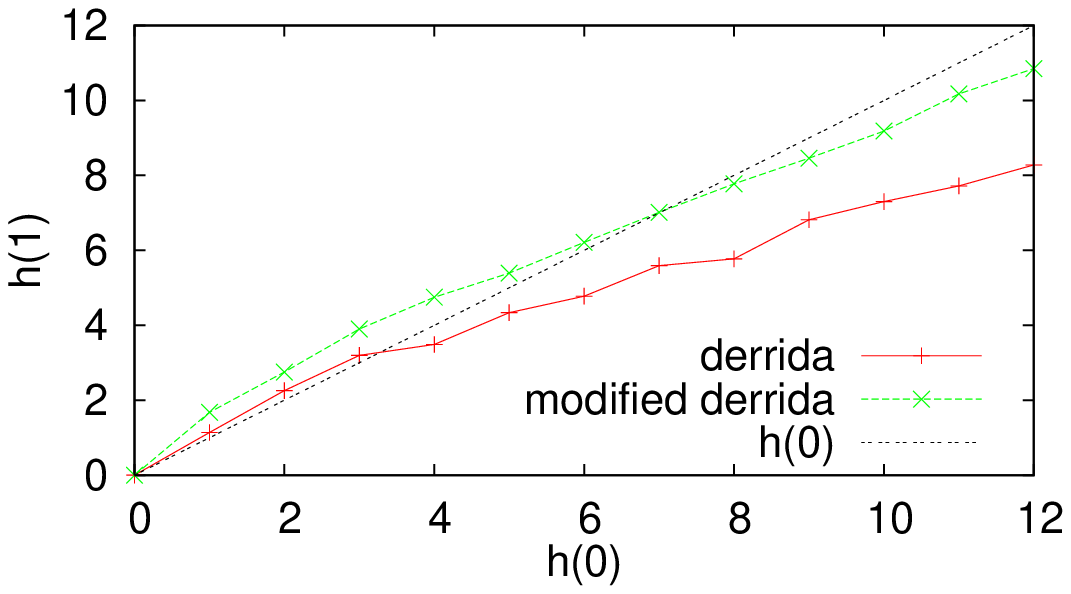}
	\caption{Generalized (top), iterated (middle, + symbols) and
          modified Derrida plot (bottom, x symbols) for the yeast
          cell-cycle network of Li et al. \cite{Li}}
	\label{yeast}
\end{figure}

Finally, we want to apply the different kinds of plots to a model of a
real biological network. The model of the yeast cell-cycle network of
Li et al. \cite{Li} is known to show a highly robust behaviour. This
robustness is due to the basin size of the main attractor rather than
to a specific state space structure \cite{Willadsen}. The model
consists of 11 key nodes, and it has Boolean update rules based on
threshold functions. A 12th node does not represent a protein, but the
cell size. Depending on whether this size is below or above a
threshold value, a new round of the cell cycle is initiated by turning
on the 12th node.  The main fixed point of the network, which attracts
more than 80 percent of all network states, corresponds to the G1
phase of the cell cycle, during which the cell grows in size. The
other attractors of the network are also fixed points.

The generalized, iterated and modified Derrida plots for this model
are shown in figure \ref{yeast}. All three graphs show again clear
deviations from the plots for RBNs. The generalized plots resemble those of
a chaotic RBNs as far as the initial slope and the almost stationary
fixed point is concerned, but for larger $h$ and higher iteration
number the slope does not become horizontal as is the case for chaotic
RBNs. This must be due to the influence of the large basin of
attraction and to long transient times which prevent the system from
showing any kind of stationary behaviour.

In contrast to the networks evolved for robustness, the two-step
Derrida plot of the yeast cell-cycle network lies above the iterated
Derrida map (figure \ref{yeast} b). This indicates that the
trajectories move away from each other faster than suggested by the
first iteration. Similarly, the modified Derrida plot lies above the
conventional one (see figure \ref{yeast} c), showing that
perturbations of attractor states have more
impact than perturbations of random states. This fits together with
the finding that the fixed point of the cell cycle is more sensitive
to perturbations than other states of the model \cite{Fretter}.

\section{Conclusion}
In this paper, we introduced different generalizations of the Derrida
plot and evaluated them for RBNs and for evolved BNs. We first showed that all the results obtained for
RBNs agree with the annealed approximation, even when networks are
small.  In particular, RBNs satisfy equation (\ref{fundamental}), that
allows us to predict the Derrida plots as long as the annealed
approximation is valid. We then evaluated these plots for networks
that were evolved for high robustness of their attractors by means of
an adaptive walk. For those networks, the annealed approximation is no
longer valid, and the temporal evolution of the Hamming distance
between two network states is no longer determined by the value of the
Hamming distance alone. Furthermore, the simulation results imply that
the standard classification of the dynamics of Boolean networks into
frozen, critical and chaotic behaviour can not be applied to evolved
networks, since these networks show features characteristic of all
three types of behaviour. Nevertheless the different kinds of Derrida
plots shown in this paper allow a more detailed insight into the state
space structure and therefore into the actual behaviour of the
networks.  We conclude that this simple classification may not be
appropriate for biological networks either, since such networks are
also shaped by their evolutionary history.  Our work thus confirms and
complements the work by Sevim and Rikvold \cite{sevim}, who arrive at
similar conclusions. Future studies of perturbation propagation in
realistic BNs will have to focus on the question of how
these networks achieve the remarkable combination of dynamical
robustness with the ability to respond flexibly to changes in the
environment.

We acknowledge partial support of this project by
  the Volkswagen Foundation and by the Deutsche Forschungsgemeinschaft
  (DFG contract no. Dr300/4-1.)

\bibliographystyle{unsrt.bst}

\end{document}